\documentstyle[12pt,fleqn]{article}
\input epsf
\def\erg{\,{\mbox{erg}}}
\def\sec{\,{\mbox{s}}}
\def\cm{\,{\mbox{cm}}}
\def\eV{\,{\mbox{eV}}}
\def\meter{\,{\mbox{m}}}
\def\km{\,{\mbox{km}}}
\def\Mpc{\,{\mbox{Mpc}}}
\def\kpc{\,{\mbox{kpc}}}
\def\rad{\,{\mbox{rad}}}

\def\be{\begin{equation}}
\def\ee{\end{equation}}
\def\ba{\begin{eqnarray}}
\def\ea{\end{eqnarray}}

\def\ga{\mathrel{\mathpalette\fun >}}
\def\fun#1#2{\lower3.6pt\vbox{\baselineskip0pt\lineskip.9pt
        \ialign{$\mathsurround=0pt#1\hfill##\hfil$\crcr#2\crcr\sim\crcr}}}
\def\pho{{{{\gamma}}}}
\def\r0{{{R^0}}}

\def\GeV{{\,{\mbox{GeV}}}}
\def\MeV{{\,{\mbox{MeV}}}}
\def\keV{{\,{\mbox{keV}}}}

\def\re#1{{[\ref{#1}]}}
\def\eqr#1{{Eq.\ (\ref{#1})}}

\begin{document}

\pagestyle{empty}

\begin{center}
{\Large \bf Are ultrahigh energy cosmic rays \\  \vspace{8pt} 
a signal for supersymmetry?}
\baselineskip=14pt
\vspace{0.75cm}

Daniel J.\ H.\ Chung\footnote{Electronic mail: 
		{\tt djchung@yukawa.uchicago.edu}}\\
{\em Department of Physics and Enrico Fermi Institute\\ 
The University of Chicago, Chicago, Illinois~~60637, and\\
NASA/Fermilab Astrophysics Center\\
Fermi National Accelerator Laboratory, Batavia, Illinois~~60510}\\
\vspace{0.4cm}
Glennys R.\ Farrar\footnote{Electronic mail: 
		{\tt farrar@farrar.rutgers.edu}}\\
{\em Department of Physics and Astronomy\\
Rutgers University, Piscataway, New Jersey~~08855}\\
\vspace{0.4cm}
Edward W.\ Kolb\footnote{Electronic mail: 
	{\tt rocky@rigoletto.fnal.gov}}\\
{\em NASA/Fermilab Astrophysics Center\\
Fermi National Accelerator Laboratory, Batavia, Illinois~~60510, and\\
Department of Astronomy and Astrophysics and Enrico Fermi Institute\\
The University of Chicago, Chicago, Illinois~~60637}\\
\end{center}
\baselineskip=24pt

\begin{quote}
\hspace*{2em}
We investigate the possibility that cosmic rays of energy larger than the
Greisen--Zatsepin--Kuzmin cutoff are not nucleons, but a new stable, massive,
hadron that appears in many extensions of the standard model.  We
focus primarily on the $S^0$, a $uds$-gluino bound state.  The range of the
$S^0$ through the cosmic background radiation is significantly longer than
the range of nucleons, and therefore can originate from sources at cosmological
distances. 
\vspace*{8pt}

PACS number(s): 96.40, 98.70, 11.30.P

\end{quote}

\renewcommand{\thefootnote}{\arabic{footnote}}
\addtocounter{footnote}{-3}

\newpage

\pagestyle{plain} 
\setcounter{page}{1}

\baselineskip=24pt
\renewcommand{\baselinestretch}{1.5}
\footnotesep=14pt

\vspace{36pt}
\centerline{\bf I. INTRODUCTION}
\vspace{24pt}

Detection of cosmic rays of energies above $10^{20}\eV$
[\ref{measure},\ref{flyseye}] has raised yet unsettled questions
regarding their origin and composition.  The first problem is that it
is difficult to imagine any astrophysical site for the cosmic
accelerator (for a review, see Ref.\ [\ref{bierman}]).  The Larmour
relation for a particle of charge $Z$, $(E/10^{18}\eV) = Z (R/\kpc)
(|\vec{B}|/\mu{\mbox{Gauss}})$, sets the scales for the required size,
$R$, and magnetic field strength, $|\vec{B}|$, of the accelerator.
One would expect any sources with sufficient $R|\vec{B}|$ to
accelerate particles to ultrahigh energies to appear quite unusual in
other regards.

A second issue is the composition of the observed cosmic rays.  The
shower profile of the highest energy event[\ref{flyseye}] is
consistent with its identification as a hadron but not as a
photon[\ref{halzen}].  Ultrahigh-energy\footnote{We use the term
ultra-high energy to mean energies beyond the GZK cutoff (discussed
below) which can be taken to be $10^{19.6}$ eV.} (UHE) events observed
in air shower arrays have muonic composition indicative of
hadrons[\ref{measure}].  The problem is that the propagation of
hadrons -- neutrons, protons, or nuclei -- over astrophysical
distances is strongly affected by the existence of the cosmic
background radiation (CBR).  Above threshold, cosmic-ray nucleons lose
energy by photoproduction of pions, $N\gamma\rightarrow N\pi$,
resulting in the Greisen--Zatsepin--Kuzmin (GZK) cutoff in the maximum
energy of cosmic-ray nucleons.  If the primary is a heavy nucleus,
then it will be photo-disintegrated by scattering with CBR photons.
Indeed, even photons of such high energies have a mean-free-path of
less than 10 Mpc due to scattering from CBR and radio
photons[\ref{elbert}].  Thus unless the primary is a neutrino, the
sources must be nearby (less than about 50 Mpc).  This would present a
severe problem, because unusual sources such as quasars and Seyfert
galaxies typically are beyond this range.

However, the primary cannot be a neutrino because the neutrino
interaction probability in the atmosphere is very small.  This would
imply an implausibly large primary flux, and worse yet, would imply
that the depths of first scattering would be uniformly distributed in
column density, contrary to observation. The suggestion that the
neutrino cross section grows to a hadronic size at UHE[\ref{halzen}]
has recently been shown to be inconsistent with unitarity and
constraints from lower energy particle physics[\ref{burdman}].

Since UHE cosmic rays should be largely unaffected by intergalactic or
galactic magnetic fields, by measuring the incident direction of the
cosmic ray it should be possible to trace back and identify the
source.  Possible candidate sources within $10^\circ$ of the UHE
cosmic ray observed by the Fly's Eye \re{flyseye} were studied in
Ref.\ \re{elbert}.\footnote{Ten degrees is taken as the extreme
possible deflection angle due to magnetic fields for a proton of this
energy.} The quasar 3C 147 and the Seyfert galaxy MCG 8-11-11 are
attractive candidates.  Lying within the $1\sigma$ error box of the
primary's incoming direction, the quasar 3C 147 has a large radio
luminosity ($7.9 \times 10^{44} \erg \sec^{-1}$) and an X-ray
luminosity of about the same order of magnitude, indicative of a large
number of strongly accelerated electrons in the region. It also
produces a large Faraday rotation, with rotation measure $\mbox{RM} =
-1510 \pm 50 \rad
\meter^{-2}$, indicative of a large magnetic field over large
distances.  It is noteworthy that this source is within the error box
of a UHE event seen by the Yakutsk detector.  However, 3C 147 lies at
a red shift of about $z=0.545$, well beyond $z < 0.0125$ adopted in
Ref.\ \re{elbert} as the distance upper limit for the source of UHE
proton primaries.  Just outside the $2 \sigma$ error box of the
primary's incoming direction is the Seyfert galaxy MCG 8-11-11.  It is
also unusual, with large X-ray and low-energy gamma-ray luminosities
($4.6
\times10^{44}\erg\sec^{-1}$ in the $20-100 \keV$ region and $7
\times10^{46} \erg\sec^{-1}$ in the $0.09-3 \MeV$ region).  At a
redshift of $z=0.0205$, it is much closer than 3C 147, but it is still
too distant for the flux to be consistent with the observed proton
flux at lower energies\re{elbert}.

Briefly stated, the problem is that there are no known candidate
astronomical sources within the range of protons, neutrons, nuclei, or
even photons.  Yet there are good candidate sources at 100-1000 Mpc.
In this paper we propose that the answer to this cosmic-ray conundrum
may be that UHE cosmic rays are not known particles but a new species
of particle we denote as the uhecron, $U$.  The meager information we
have about the cosmic ray events allows us to assemble a profile for
the properties of the uhecron:

1) The uhecron interacts strongly: Although there are only a handful
of UHE events, the observed shower development and muonic content suggests
a strongly interacting primary.

2) The uhecron is stable or very long lived: Clearly if the particle
originates from cosmological distance, it must be stable, or at least
remarkably long lived, with $\tau \ga 10^6{\rm s}(m_U/3\,\mbox{GeV})
(L/1\, \mbox{Gpc})$ where $L$ is the distance to the source.

3) The uhecron is massive, with mass greater than about 2 GeV: If the
cosmic ray is massive, the threshold energy for pion production
increases, and the energy lost per scattering on a CBR photon will
decrease.  We will go into the details of energy loss later in the paper,
but this general feature can be understood from simple kinematics.  In
$U \gamma 
\rightarrow U\pi$, the threshold for pion production is $s_{\rm min} =
m_U^2 + m_\pi^2 + 2m_U m_\pi$.  In the cosmic-ray frame where the $U$
has energy $E_U \gg m_U$ and the photon has energy $E_\gamma \sim 3T $
(where $T = 2.4\times10^{-4}\eV$ is the temperature of the CBR), $s
\simeq m_U^2 + 4 E_\gamma E_U$.  Thus, the threshold for pion
production, $s\geq s_{\rm min}$, results in the limit $E_U\ga m_\pi
m_U/(2 E_\gamma)$.  More generally, the threshold for producing a
resonance of mass $M_R = M_U + \Delta$ is $E_U = \Delta m_U/(2
E_\gamma)$. For $E_\gamma=3T$, and if the uhecron is the proton, the
threshold for pion photoproduction is $E_U \approx 10^{20}\eV$.  Of
course the actual threshold is more involved because there is a
distribution in photon energy and scattering angle, but the obvious
lesson is that if the mass of the primary is increased, the threshold
for pion production increases, and the corresponding GZK cutoff will
increase with the mass of the cosmic ray.  Furthermore, since the
fractional energy loss will be of order $ m_\pi/m_U$, a massive
uhecron will lose energy via pion-photoproduction at a slower rate
than a lighter particle.  Another potential bonus if the cosmic ray is
not a neutron or a proton is that the cross section for $U \gamma
\rightarrow U \pi$ near threshold may not be strongly enhanced by a
resonance such as $\Delta(1232)$, as when the $U$ is a nucleon.
Although there may well be a resonance in the $U \pi$ channel, it
might not have the strength or be as near the pion-photoproduction
threshold as the $\Delta(1232)$ is in the pion-nucleon channel.

4) We will assume the uhecron is electrically neutral: Although not as
crucial a requirement as the first three, there are three advantages
if the uhecron is neutral.  The first is that it will not lose energy
through $e^+e^-$ pair production off the CBR photons.  Another
advantage of a neutral particle is that because it will be unaffected
by intergalactic and galactic magnetic fields, its arrival direction
on the sky will point back to its source.  Thirdly, there will be no
energy losses due to synchrotron or bremsstrahlung radiation.  Of
course because a neutral particles will not be accelerated by normal
electromagnetic mechanisms, it is necessary to provide at least a
plausibility argument that they can be produced near the source.  For
instance, they may be produced as secondaries in collisions induced by
high-energy protons.

In this paper we analyze the possibility that a supersymmetric baryon
$S^0$ ($uds$-gluino bound state whose mass is 1.9-2.3 GeV -- see
below) is the uhecron instead of the proton, as first proposed in
Ref.\ [\ref{gf96}].  The $S^0$ has strong interactions, it can be
stable, it is more massive than the nucleon, and it is neutral with
vanishing magnetic moment [\ref{gf96}].   Remarkably, this particle is
not experimentally excluded.  The light gluino required in this scenario
would have escaped detection.   Experimental limits and signatures are
discussed in [\ref{gf96}] and the reviews of Farrar [\ref{gf97},
\ref{farrarprd95}].

If UHE cosmic rays are $S^0$s, we will show that their range is at least
an order of magnitude greater than that of a proton, putting MCG 8-11-11
(and possibly even 3C 147) within range of the Fly's Eye event. 

While the main thrust of this paper is an investigation into the
scenario where the $S^0$ is the uhecron, most of our analysis can also
be applied to the case where the uhecron is much more massive than
assumed for the $S^0$.  Extensions of the standard model often predict
new heavy, e.g., multi-TeV, colored particles which in some instances
have a conserved or almost-conserved quantum number.  Bound to light
quarks these form heavy hadrons, the lightest of which would be stable
or quasistable.  Such a particle would propagate through the CBR
without significant energy loss because the threshold energy for
inelastic collisions is proportional to its mass.  Some mechanisms for
uhecron production discussed below would be applicable for a new very
massive hadron.  However such a particle probably would not be an
acceptable candidate for the uhecron because its interaction in the
atmosphere is quite different from that of nucleons, nuclei, or an
$S^0$.  Although it is strongly interacting, its fractional energy
loss per collision in the Earth's atmosphere is only of order
$(1\,{\rm GeV}/M)$, where $M$ is the mass of the heavy
hadron.\footnote{In the infinite momentum frame for the heavy hadron,
this is the fractional momentum carried by light partons since they
have the same velocity as the heavy parton, but their mass is of order
$\Lambda_{QCD}$.  It is the momentum of these light partons which is
redistributed in a hadronic collision. Of course a hard collision with
the heavy quark would produce a large fractional energy loss, but the
cross section for such a collision is small: $\approx
\alpha_s^2/E^2$.}  Thus if the uhecron energy deposition spectrum is
indeed typical of a nucleon or nucleus, as present evidence suggests,
we cannot identify the uhecron with a very massive stable hadron.  The
maximum uhecron mass consistent with observed shower properties is
presently under investigation[\ref{afk}].

\vspace{36pt}
\centerline{\bf II.  PRODUCTION OF UHE $S^0$s}
\vspace{24pt}

We first address the question of whether there is a plausible scenario
to produce UHE $S^0$s.  This is a tricky question, since there is no
clear consensus on the acceleration mechanism even if the primary
particle is a proton.  Here we simply assume that somehow UHE protons
are produced, and ask if there is some way to turn UHE protons into UHE
$S^0$s.  Our intent is not to establish the viability of any particular mechanism
but to see that finding a satisfactory mechanism is not dramatically more
difficult than it is for protons. 

Assuming that there exists an astrophysical accelerator that can
accelerate protons to energies above $10^{21}\eV$, one can envisage a
plausible scenario of $S^0$ production through proton collision with
hadronic matter surrounding the accelerator.  A $p$-nucleon collision
will result in the production of $R_p$'s, the $uud$-gluino state whose
mass is about 200 MeV above the $S^0$.  The $R_p$ decays to an $S^0$
and a $\pi^+$,\footnote{The decay $R_p \rightarrow S^0 \pi$ was the
subject of an experimental search \re{albuquerque}.  However the
sensitivity was insufficient in the mass and lifetime range of
interest ($m(R_p) = 2.1 - 2.5$ GeV, $\tau(R_p) = 2 \cdot 10^{-10} - 2
\cdot 10^{-11} $ sec \re{gf96}) for a signal to have been expected.}
with the $S^0$ receiving a momentum fraction of about
$(m_{S^0}/m_{R_p})^2$.  From a triple Regge model of the collision,
one estimates that the distribution of the produced $R_p$'s as a
function of the outgoing momentum fraction $x$ is $d \sigma/d x \sim
(1-x)^{1 - 2 \alpha} (s')^{\alpha_P - 1}$ as $x$ approaches unity.
Here, $s' = (1-x)s$ and $\alpha$ is the Regge intercept of the
SUSY-partner of the Pomeron.  Thus, $\alpha = \alpha_P - 1/2 =
\epsilon + 1/2$, where $\epsilon \approx 0.1$ is the amount the
pomeron trajectory is above 1 at high energies.  Hence, we
parameterize the $S^0$ production cross section in a $p$-nucleon
collision as $d\sigma/dx = A E_p^\epsilon$; $x$ is the ratio of the
$S^0$ energy to the incident energy.  Parameterizing the high energy
proton flux from the cosmic accelerator as $dN_p/dE_p = B
E_p^{-\gamma}$, we have a final $S^0$ flux of $dN_{S^0}/dE = \kappa n
L A B E^{-\gamma+\epsilon}$, where $n L$ is the matter column density
with which the proton interacts to produce an $R_p$ and $\kappa$ is of
order 1 (for $\gamma = 2$, $\kappa = 0.4$).  Note that the produced
$S^0$'s are distributed according to a spectrum that is a bit flatter
than the high energy proton spectrum.

A disadvantage to this ``beam-dump'' $S^0$ production mechanism is the
suppression factor of about $A E^\epsilon/\sigma_{p p}$, where
$\sigma_{p p}$ is the proton-proton total diffractive cross
section.  This suppression could be of order
$10^{-1}$ to $10^{-2}$ for typical energies.  However the
produced $S^0$'s enjoy a compensating advantage.  The large column 
densities characteristic of most candidate acceleration regions makes
it hard to avoid energy degradation of protons before they escape.
That is, $ L ( n_p \sigma_{p N} + n_e \sigma_{p e} + n_\gamma
\sigma_{p \gamma})$ may be much greater than unity.  By contrast, $S^0$'s
may escape with little or no energy loss.  Their electromagnetic
interactions are negligible, and analogy with glueball wavefunctions
suggests that $\sigma_{S^0 N}$ could be as small as $10^{-1}\sigma_{pN}$
\re{gf96}. 
Thus the emerging $S^0$ and nucleon fluxes could be of the same order of
magnitude.  This would be necessary for a very distant source such
as 3C 147 to be acceptable, since the required particle flux for the 
detected flux on Earth already pushes its luminosity limit.  Assuming 
that the $3.2 \times 10^{20}\eV$ event of Fly's Eye came during its
exposure to 3C 147, the resulting time-averaged flux is $11
\eV\cm^{-2}\sec^{-1}$, which is greater than the X-ray luminosity of
3C 147 \re{elbert}.  

In connection with the ``beam dump" mechanism, we note that it is
possible to have at the source a nucleon flux significantly greater
than the $S^0$ flux, and yet at Earth still have a large enough $S^0$
flux to account for the high energy end of the spectrum without being
inconsistent with the rest of the observed cosmic ray spectrum.
To see this is possible, suppose as an illustrative example that the $S^0$
spectrum for energies above $10^{20}{\rm eV}$ is a smooth extrapolation
of the proton spectrum at energies below the GZK cutoff, i.e., if $J_{p}(E)
= A E^{-3}$ for $E < 10^{19.6}$ eV then $J_{S^0}(E) = A E^{-3}$ for $E >
10^{20}$ eV.  Denoting the $S^0$--to--proton suppression factor by $\eta$,
the proton flux for $E > 10^{20}$ eV is then $J_{p}(E) = \eta^{-1} A E^{-3}$.
Protons of energy greater than the GZK cutoff (here taken to be $10^{19.6}$eV)
will bunch up in the decade in energy below the GZK cutoff [\ref{hs},\ref{gqw}].
The total number in the pile-up region will receive a contribution from protons
from the source above the GZK cutoff as well as those originally in the pile-up
region.  With $\eta=10^{-2}$, there will be equal contributions from the
pile-up protons and the protons originally below the source.  The statistics
of the number of events with energy above $10^{18.5}$eV are too poor
to exclude this scenario; indeed there is some indication of a bump in
the spectrum in this region[\ref{measure}].

Note that even for a point source as far away as 1200 Mpc (e.g. 3C
147), the required flux of high energy protons at the accelerator is
not unacceptable.  For instance extrapolating the spectrum as 
$7.36\times10^{18} E^{-2.7} ~{\rm eV/m^2/sr/sec}$ and using our pessimistic
efficiency for $S^0$ production (factor of 1/100), requires the high energy
proton luminosity of the source to be $\sim 10^{47}$ ergs/sec.  This
is indeed a high value, but not impossible.

Another possible mechanism of high energy $S^0$ production is the
direct acceleration of charged light SUSY hadrons (mass around
$2-3\GeV$), such as $R_p$ and $R_\Omega$, whose lifetime is about
$2 \cdot 10^{-10} - 2 \cdot 10^{-11} $ sec \re{gf96}.  Due to the
large time-dilation factor ($E/m\approx 10^{11}$), whatever
electromagnetic mechanism accelerates the protons may also be able to
accelerate the high energy SUSY hadrons.  Then, one can imagine that
the high energy tail of the hadronic plasma which gets accelerated by
some electromagnetic mechanism will consist of a statistical mixture
of all light strong-interaction-stable charged hadrons.  In that case
the flux of the resulting $S^0$ will have the same spectrum as the
protons, differing in magnitude by a factor of order unity, which
depends on the amount of SUSY hadrons making up the statistical
mixture.  Conventional shock wave acceleration mechanisms probably
require a too long time scale for this mechanism to be feasible (e.g.,
Ref.\ \re{rachen}).  However, some electromagnetic ``one push''
mechanisms similar to the one involving electric fields around
pulsars [\ref{onepush}] may allow this kind of acceleration if the
electric field can be large enough.  It is certainly tantalizing that
the time scale of the short time structure of pulsars and gamma ray
bursts is consistent with the scale implied by the time-dilated
lifetime of charged $R$-baryons.  

A somewhat remote possibility is that there may be gravitational
acceleration mechanisms which would not work for a charged particle
(because of radiation energy losses and magnetic confinement) but
would work for a neutral, zero magnetic moment particle such as an
$S^0$.  For example, if $S^0$'s exist in the high energy tail of the
distribution of accreting mass near a black hole (either by being
gravitationally pulled in themselves or by being produced by a proton
collision), they may be able to escape with a large energy.
 A charged particle, on the other
hand, will not be able to escape due to radiation losses.
Unfortunately, this scenario may run into low flux problems due to its
reliance on the tail of an energy distribution.

A final possibility is the decay of a long-lived superheavy relics of
the big bang, which  would produce all light particles present in the
low-energy world, including the $S^0$.  For instance if such relics
decay via quarks which then fragment, as in models such as Ref.\
\re{kr}, the $S^0$/nucleon ratio is probably in the range $10^{-1} $
to $10^{-2}$ based on a factor of about $10$ suppression in producing
a 4-constituent rather than 3-constitutent object, and possibly some
additional suppression due to the $S^0$'s higher mass.\footnote{After
our work was completed, Ref.\ \re{berezinsky} appeared with an
estimate of the production of gluino-hadrons from the decay of cosmic
necklaces.  Note that their pessimism regarding the light gluino
scenario is mostly based on arguments which have been rebutted in the
literature (see for example Refs.\ \re{plaga} and \re{farrarprd95}).}

Of the scenarios considered above, only the last two are conceivably
relevant for a super-heavy (0.1 - 1000 TeV) uhecron.  Although the
energy in $p$--nucleon collisions ($\sqrt{s} = \sqrt{2E_p m_p} \sim
10^3$TeV for primary proton energy of $10^{21}$eV) is sufficient for
superheavy particle production, the production cross section is too
small for the ``beam dump" mechanism to be efficient.\footnote{ The
cross section is proportional to the initial parton density at $x \sim
M_U/\sqrt{s}$ times the parton-level cross section, which scales as
$M_U^{-2}$.} Also, the direct acceleration mechanism is not useful for
a superheavy uhecron unless it is itself charged or is produced in the
decay of a sufficiently long-lived charged progenitor.  Even if a
sufficient density of superheavy hadrons could be generated in spite
of the small production cross section, the time scale required for the
early stages of acceleration could be too long since it is
proportional to $\beta^2$.  This leaves the decay of a superheavy
relic (either a particle or cosmic defect) as the most promising
source of uhecrons if their masses are greater than tens of GeV.

\vspace{36pt}
\centerline{\bf III.  PROPAGATION OF UHE COSMIC RAYS}
\vspace{24pt}

To calculate the energy loss due to the primary's interaction with the
CBR, we follow the continuous, mean energy loss approximation used in
Refs.\ \re{gqw} and \re{jwc}.  In this approximation we smooth over the discrete
nature of the scattering processes, neglecting the stochastic nature
of the energy loss, to write a continuous differential equation for
the time evolution of the primary energy of a single particle.  The
proper interpretation of our result is the mean energy of an ensemble
of primaries traveling through the CBR.  We shall now delineate the
construction of the differential equation.

For an ultrahigh energy proton (near $10^{20}\eV$ in CBR
frame\footnote{Let this be the frame in which CBR has a isotropic
distribution.}),  three main mechanisms contribute to the depletion of
the particle's energy: pion-photoproduction, $e^+ e^-$ pair
production, and the cosmological redshift of the momentum.
Pion-photoproduction consists of the reactions $p \gamma \rightarrow
\pi^0 p$ and $p \gamma \rightarrow \pi^+ n$.  Pion-photoproduction,
which proceeds by excitation of a resonance, is the strongest source
of energy loss for energies above about $10^{20}\eV$, while below
about $10^{19.5}\eV$, $e^+ e^-$ pair production dominates.  For the
scattering processes (pion-photoproduction and $e^+ e^-$ pair
production), the mean change in the proton energy ($E_p$) per unit
time (in the CBR frame) is 
\begin{equation} 
\frac{dE_p(\mbox{scatter})}{dt} = -
\sum_{\mbox{events}} (\mbox{mean event rate}) \times \Delta E
\label{eq:lossrate}
\end{equation}
where the sum is over distinct scattering events with an energy loss
of $\Delta E$ per event.  The mean event rate is given by \be
\mbox{mean event rate}= \frac{1}{\gamma} \frac{d \sigma}{d \xi}
f(E_\gamma) dE_\gamma d\xi \ee where $\gamma=E_p/m_p$ is necessary to
convert from the event rate in the proton frame (proton's rest frame),
where we perform the calculation, to the CBR frame, $d\sigma/d\xi$ is
the differential cross section in the proton frame,\footnote{The
differential $d \xi$ is $dQ d\eta$ ($Q$ and $\eta$ are defined below)
for the $e^+ e^-$ pair production while it is $d \cos \theta$ for the
pion-photoproduction.} and $f$ is the number of photons per energy per
volume in the proton frame.  To obtain $f$ we start with the isotropic
Planck distribution and then boost it with the velocity parameter
$\beta$ to the proton frame 
\begin{equation}
n(E_\gamma, \theta)=\frac{1}{(2
\pi)^3} \left[\frac{2 E_\gamma^2}{\exp[\gamma E_\gamma (1+ \beta \cos
\theta)/T]-1} \right]
\label{eq:protonplanck}
\end{equation}
where $\theta$ is the angle that the photon direction makes with
respect to the boost direction.  Integrating \eqr{eq:protonplanck}
over the solid angle\footnote{The exact angular integration range is
unimportant as long as the range encompasses $\cos \theta = -1$ (where
the photon distribution is strongly peaked in the ultrarelativistic
limit) since we will be taking the ultrarelativistic limit.} and
taking the ultrarelativistic limit, we find
\be
f=\frac{E_\gamma T}{2 \pi^2 \gamma} \ln\left[ \frac{1}{1-\exp(-E_\gamma/2
\gamma T)} \right].
\label{eq:phonum}
\ee
For $\Delta E$, the energy loss per event in the
CBR frame, we can write
\be
\Delta E(\cos \theta, p_r) =\gamma m_p \left[ 1 +\frac{\beta
p_r}{m_p} \cos \theta - \sqrt{1+ \left(\frac{p_r}{m_p} \right)^2} \ \right]
\ee
where $p_r$, which may depend on $E_\gamma$ and $\cos \theta$, is the
recoil momentum of the proton and $\theta$ is the 
angle between the incoming photon direction and the outgoing proton
direction.  Putting all these together, the energy loss rate due to
scattering given by \eqr{eq:lossrate} becomes
\begin{eqnarray}
 \frac{dE_p(\mbox{scatter})}{dt} & = &  
-\gamma^{-1} \int dE_\gamma f(E_\gamma) \cr
& & \times\sum_i \int 
d\xi_i \frac{d \sigma_i}{d \xi_i} (E_\gamma, \xi_i)\,  \Delta E
(\cos \theta(E_\gamma, \xi_i), p_r(E_\gamma, \xi_i)) 
\label{eq:scatterdepdt}
\end{eqnarray}
where only functions yet to be specified are the recoil momentum
and the differential cross section (for each type of reaction $i$).

For the reaction involving the production of a single pion, the recoil
momentum of the protons in the proton frame can be expressed as
\begin{eqnarray}
p_r(E_\gamma, \cos \theta)  & = & \cr
& & \hspace*{-3em} \frac{2 q^2 E_\gamma \cos \theta \pm
(E_\gamma + m_p) 
\sqrt{4 E_\gamma^2 m_p^2 
\cos^2 \theta - 4 m_\pi^2 m_p (E_\gamma+m_p)+m_\pi^4}}{2\, 
[ (E_\gamma + m_p)^2 -
E_\gamma^2 \cos^2 \theta]}
\label{eq:pr}
\end{eqnarray}
where $q^2=m_p (m_p+E_\gamma) - m_\pi^2/2$.  When the photon energy
$E_\gamma$ is approximately at threshold energy of $m_\pi + m_\pi^2/2
m_p$ and the proton recoils in the direction $\theta=0$, the recoil
momentum is about $m_\pi$.  The recoil momentum is a double valued
function, where the negative branch corresponds to the situation where
most of the photon's incoming momentum is absorbed by the pion going
out in the direction of the incoming photon.  Thus, since the positive
branch will be more effective in retarding the proton (in the CBR
frame), we will neglect the negative branch to obtain a conservative
estimate of the ``cutoff'' distance.  It is possible to work out the
kinematics for multipion production, but for our purpose of making a
reasonably conservative estimate, it is adequate to use
\eqr{eq:pr} as the recoil momentum even for multipion
production.\footnote{For example, one can easily verify that the
maximum proton recoil during one pion production is greater than the
maximum proton recoil during two pion production.}

The pion-photoproduction cross section has been estimated by assuming
that the $s$-wave contribution dominates, which would certainly be true
near the threshold of the production.  The cross section is taken to
be a sum of a Breit-Wigner piece and two non-resonant pieces:
\begin{eqnarray}
\sigma(\mbox{pion}) & = & 2 \sigma_{1\pi} \ \Theta\! \left (E_\gamma - 
m_\pi - \frac{m_\pi^2}{2 m_p}\right) +
 2 \sigma_{\mbox{multipion}} \nonumber \\
\sigma_{1\pi} & = & 
\frac{4 \pi}{p_{cm}^2} \left[ \frac{m_\Delta^2 \Gamma(\Delta
\rightarrow \gamma p) \Gamma(\Delta \rightarrow \pi
P)}{(m_\Delta^2-s)^2 + m_\Delta^2 \Gamma_{\mbox{tot}}^2} \right] +
\sigma_{\mbox{nonres}} \nonumber \\
\Gamma(\Delta \rightarrow X p) & = & \frac{p_{cm}^X \omega_X}{8
m_\Delta \sqrt{s}} \nonumber \\
\Gamma_{\mbox{tot}} & = & \frac{p_{\mbox{cm}}^\pi}{\sqrt{s}} 
\frac{ 2 m_\Delta^2 \overline{\Gamma}_{\mbox{tot}}}{\sqrt{[m_\Delta^2 -
(m_\pi+m_p)^2]\, [m_\Delta^2-(m_p-m_\pi)^2]}} \nonumber \\
\sigma_{\mbox{nonres}} & = & \frac{1}{16 \pi s}
\frac{\sqrt{[s-(m_p+m_\pi)^2] \, [s-(m_p-m_\pi)^2]}}{(s-m_p^2)} |{\cal
M}(p \gamma \rightarrow \pi p)|^2 \nonumber \\
\sigma_{\mbox{multipion}} & = & a \tanh\left( \frac{E_\gamma -
E_{\mbox{multi}}}{m_\pi}\right) \Theta(E_\gamma - E_{\mbox{multi}}) 
\end{eqnarray}
where $\omega_X$ is defined through $4 \pi \omega_X \equiv \int d\Omega |
{\cal M}(\Delta \rightarrow X p) |^2$, ${\cal M}$ denotes an
invariant amplitude, the center of momentum momentum is given as usual
by 
\begin{equation}
p_{cm}^X=\sqrt{\frac{[s-(m_p+m_X)^2]\, [s-(m_p-m_X)^2] }{4s}},
\end{equation}
and $\sigma_{\mbox{multipion}}$ is a crude approximation\footnote{The
functional form was chosen  to account for the shape
of the cross section given in Ref.\ \re{crossdata}.} for the
contribution from the multipion production whose threshold is at
$E_{\mbox{multi}} = 2 (m_\pi +m_\pi^2/m_p)$.  The $\sigma_\pi$
component of the cross section is fit\footnote{The fit is
qualitatively good, but only tolerable quantitatively.  The fit to the
data in the range between $0.212 \GeV$ and $0.4 \GeV$ resulted in a reduced
$\chi_{16}^2 \sim 50$ (due to relatively small error bars).  This
is sufficient for our purposes since our results should depend
mainly upon the gross features of the cross section.} to the $p \pho
\rightarrow n \pi^0$ data of Ref. \re{crossdata}, while the amplitude
$a$ for $\sigma_{\mbox{multipion}}$ is estimated from the $p \pho
\rightarrow X p$ data for energies $E_\gamma \ga 0.6\GeV$.  The
numerical values of the parameters resulting from the fit are
$(\omega_\gamma \omega_\pi) = 0.086\GeV^4, |{\cal M}(p\gamma
\rightarrow \pi p)|=0.018$,
$\overline{\Gamma}_{\mbox{tot}}=0.111\GeV$, $m_\Delta=1.23\GeV$, and
$a=0.2$mb.  The factor of  2 multiplying $\sigma_\pi$
accounts for the two reactions $p \gamma \rightarrow \pi^0 p$ and $p
\gamma \rightarrow \pi^+ n$, since a neutron behaves, to first
approximation, just like the proton.  
For example, the dominant pion-photoproduction reactions
involving neutrons are $n \gamma \rightarrow \pi^0 n$ and $n \gamma
\rightarrow \pi^- p$ which have similar cross sections as the
analogous equations for protons.  Thus,  we are really
estimating the energy loss of a nucleon, and not just a proton.

Taking the $p \gamma \rightarrow e^+ e^- p$ differential cross section
from Ref.\ \re{jost} (as done in Ref. \re{gqw}), we use\footnote{We
ignore that $n$ does not pair produce $e^+ e^-$.  However, this has
consequences only for energies below about $10^{19.5}$ eV.}
\begin{eqnarray}
\frac{d \sigma( \mbox{pair})}{dQ d\eta} & = & \Theta(E_\pho-2 m_e) \times
\frac{4 \alpha^3 }{E_\gamma^2}\frac{1}{Q^2} \left\{ \ln \left(
\frac{1-w}{1+w} \right) \left[ \left(1- \frac{E_\gamma^2}{4 m_e^2
\eta^2}\right) \right. \right. \cr
& &\hspace{-5em} \left.  \times \left( 1- \frac{1}{4 \eta^2} + 
\frac{1}{2\eta Q} - 
\frac{1}{8 Q^2 \eta^2} - \frac{Q}{\eta} + \frac{Q^2}{2 \eta^2} \right)
+ \frac{E_\gamma^2}{8 m_e^2 \eta^4} \right] 
\nonumber
\\ & & \hspace{-5em} \left. +
w \left[ \left(1 - \frac{E_\gamma^2}{4 m_e^2 \eta^2} \right) \left(
1- \frac{1}{4 \eta^2} + \frac{1}{2\eta Q} \right) + \frac{1}{\eta^2}
\left( 1 - \frac{ E_\gamma^2}{2 m_e^2 \eta^2} \right) (-2 Q \eta +
Q^2) \right] \right\} ,
\end{eqnarray}
where $w =\left[1-1/(2Q \eta - Q^2)\right]^{1/2}$.  The recoil momentum
is contained in $Q=p_r/2 m_e$, and the photon energy is contained in
$\eta=E_\gamma \cos \theta /2 m_e$.  

The final ingredient in our energy loss formula is the redshift due to
Hubble expansion.  We assume a matter-dominated, flat FRW universe
with no cosmological constant.  Thus, the cosmological scale factor is
proportional to $t^{2/3}$.  The energy loss for relativistic
particles (such as our high energy proton) due to redshift is then
given by
\be
\frac{dE_p(\mbox{redshift})}{dt}= - \frac{2 E_p}{3 t}.
\label{eq:redshift}
\ee
Furthermore, note that the expansion of the universe causes the
temperature to vary with time as $t^{-2/3}$.

Adding Eqs.\ (\ref{eq:scatterdepdt}) and (\ref{eq:redshift}), we have
the proton energy loss equation 
\be 
\frac{dE_p}{dt}=
\frac{dE_p(\mbox{scatter})}{dt}+\frac{dE_p(\mbox{redshift})}{dt} \ ,
\ee
whose integration from some initial cosmological time $t_i$ to the
present time $t_0$ gives the present energy of the proton that was
injected with energy $E_i$ at time $t_i$.  Note that we are interested
in plotting $E_p(t_0)$ as a function of $t_0-t_i$ with $t_0$ fixed,
which is not equivalent to fixing $t_i$ and varying $t_0$ because there
is no time translational invariance in a FRW universe.  Note also
that we need to set the Hubble parameter $h$ (where the Hubble
constant is $100 h \km\sec^{-1}\Mpc^{-1}$) in our calculation because the
conversion between time and the redshift depends on $h$.  To show the
degree of sensitivity of our results to $h$ we will calculate the
energy loss for $h=0.5$ and $h=0.8$.

Now, suppose the primary cosmic ray is an $S^0$ instead of a proton.
The $e^+ e^-$ pair production will be absent (to the level of our
approximation) because of the neutrality of $S^0$.  Furthermore, the
mass splitting between $S^0$ and any one of the nearby resonances that
can be excited in a $\pho S^0$ interaction is larger than the
proton-$\Delta$ mass splitting, leading to a further increase in the
attenuation length of the primary.  Perhaps most importantly, the mass
of $S^0$ being about two times that of the proton increases the
attenuation length significantly because of two effects.  One obvious
effect is seen in
\eqr{eq:pr}, where the fractional energy loss per collision to leading
approximation is proportional to $p_r/m_p$ while $p_r$ has a maximum
value of about $m_\pi$.  Replacement of $m_p\rightarrow m_{S^0}$
obviously leads to a smaller energy loss per collision.
The second effect is seen in Eqs.\
(\ref{eq:phonum}) and (\ref{eq:scatterdepdt}), where for the bulk of
the photon energy integration region, a decrease in $\gamma$ (in the
exponent) resulting from an increase in the primary's mass suppresses
the photon number.  In fact, it is easy to show that if we treat the
cross section to be a constant, the pion-photoproduction contribution
to the right hand side of \eqr{eq:scatterdepdt} can be roughly
approximated as
\be
\frac{dE_p(\pi)}{dt} \approx - \frac{m_\pi^2 T^2 \sigma}{\pi^2} \exp(-y/2)
\left( 1+ \frac{3}{y} + \frac{4}{y^2} \right)
\ee
where $y=m_\pi m_p/(E_p T)$, clearly showing a significant increase in
the attenuation length as $m_p$ is replaced by $m_{S^0}$.

\begin{figure}[p]
\hspace*{25pt} \epsfxsize=400pt \epsfbox{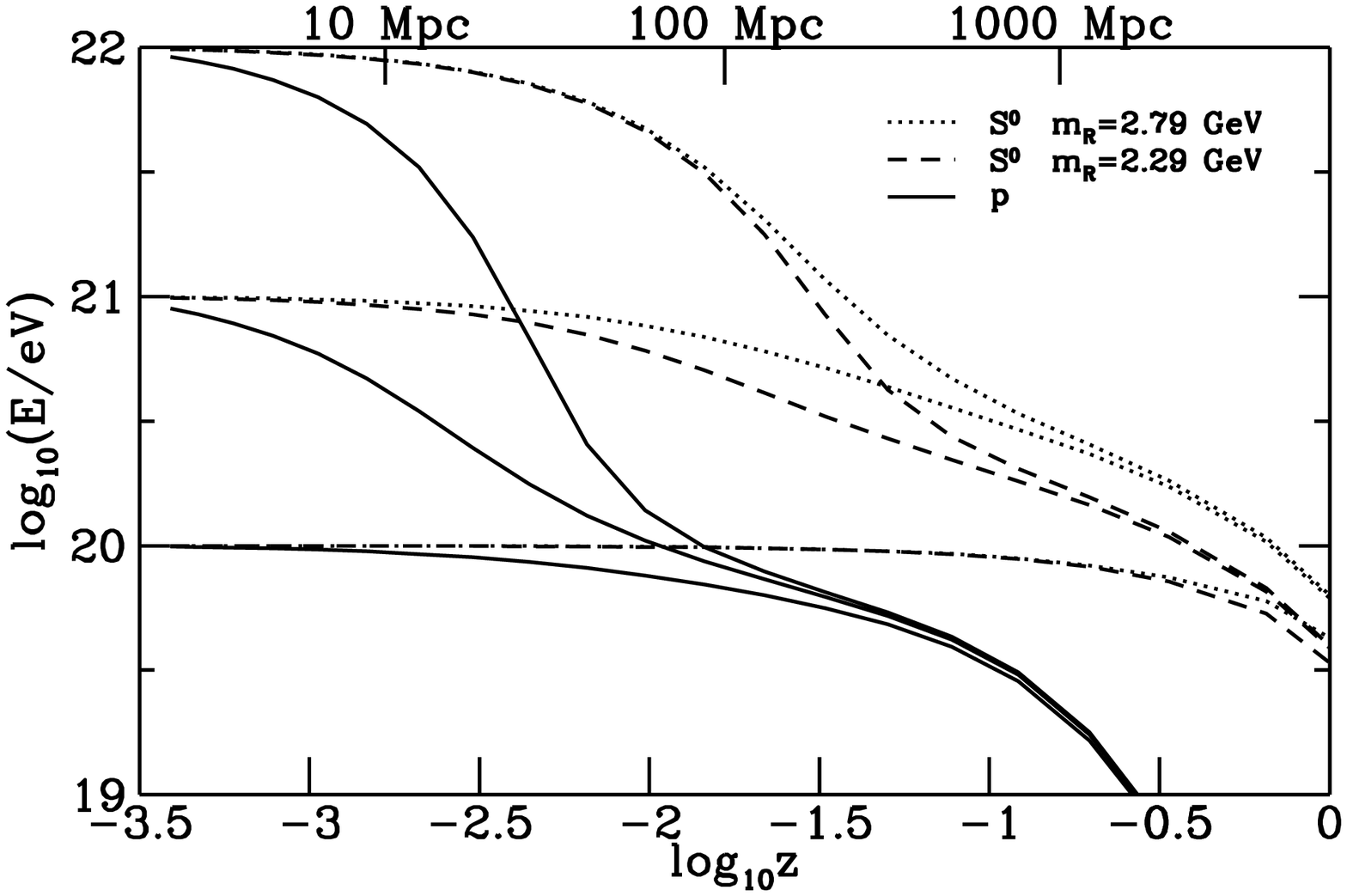}\\
\hspace*{25pt} \epsfxsize=400pt \epsfbox{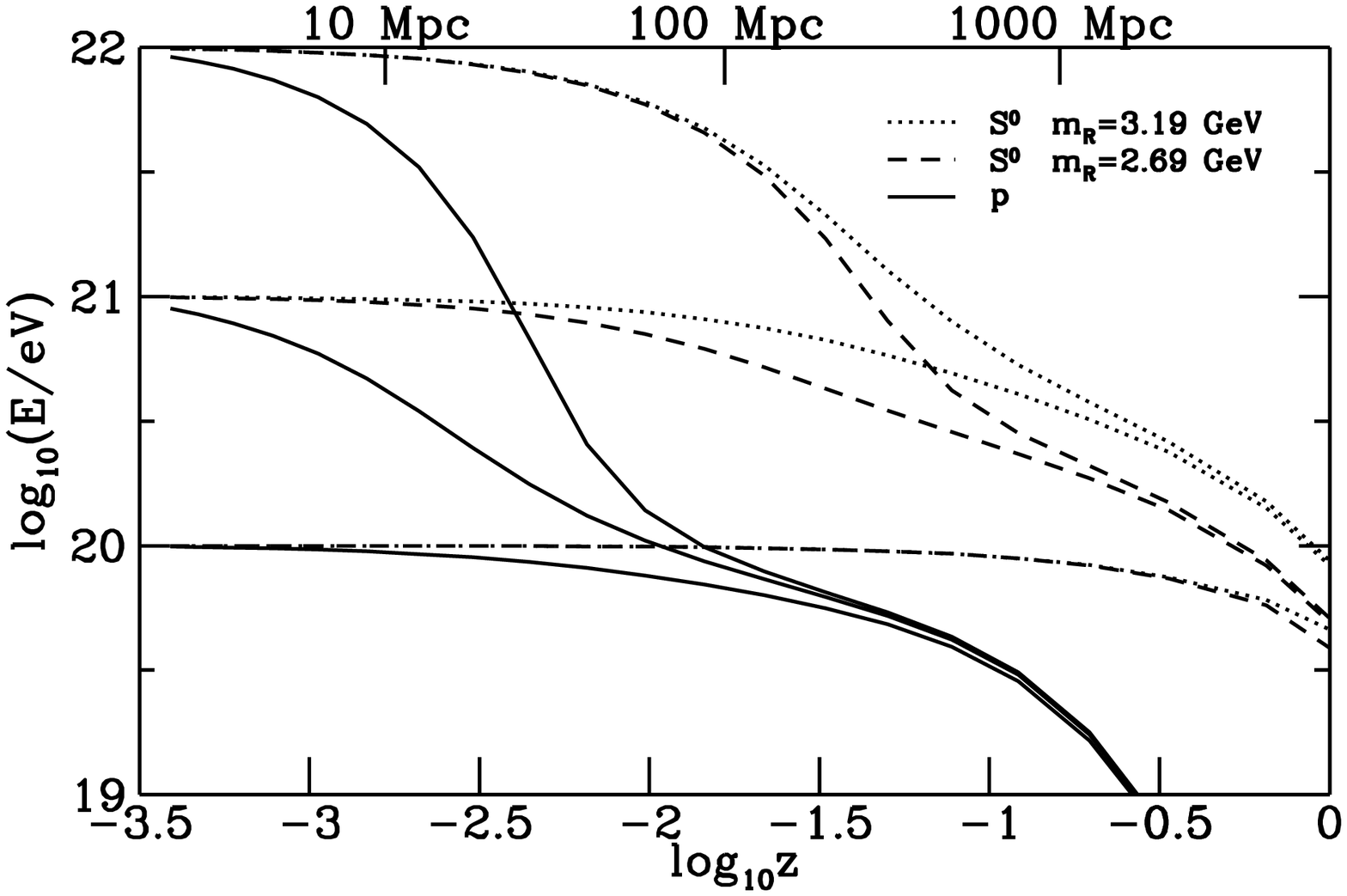}\\
\hspace*{1em} {\footnotesize{{\bf Fig.\ 1:} The figures show the
primary particle's energy as it would be observed on Earth today if it
were injected with various energies ($10^{22}\eV$, $10^{21}\eV$, and
$10^{20}\eV$) at various redshifts.  The distances correspond to
luminosity distances.  The mass of $S^0$ is $1.9 \GeV$ in the upper plot
while it is $2.3 \GeV$ in the lower plot.  
 Here, the Hubble constant has been set to $50
\km\sec^{-1}\Mpc^{-1}$. }}
\end{figure}

\begin{figure}[p]
\hspace*{25pt} \epsfxsize=400pt \epsfbox{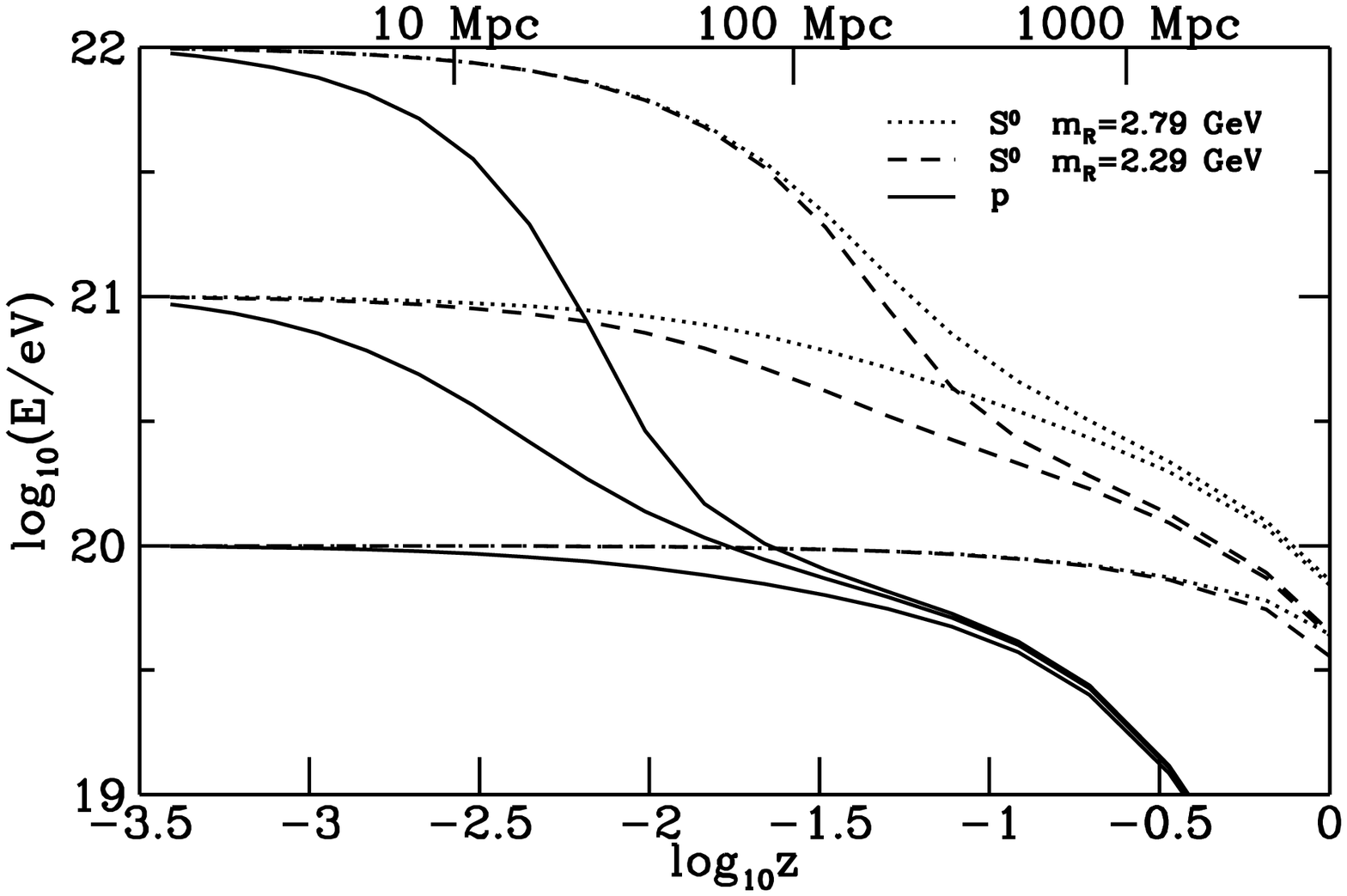}\\
\hspace*{25pt} \epsfxsize=400pt \epsfbox{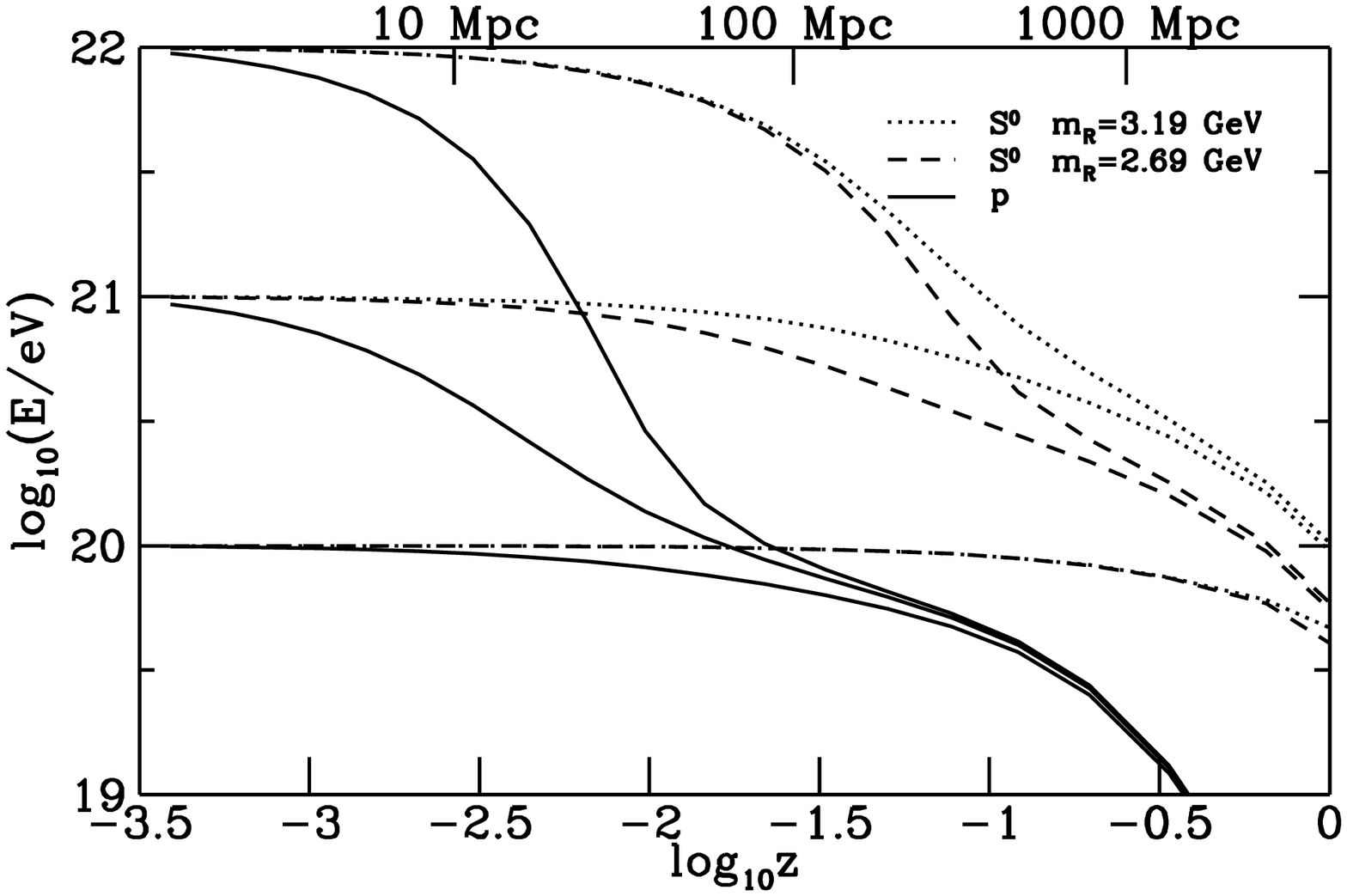}\\
\hspace*{1em} {\footnotesize{{\bf Fig.\ 2:}  Same as Fig.\ 1 except
with the Hubble constant equal to $80 \km\sec^{-1}\Mpc^{-1}$.}}
\end{figure}

The relevant resonances for the $S^0 \gamma$ collisions are spin-1
$R_\Lambda$ and $R_\Sigma$ [\ref{gf96}] (whose constituents are those
of the usual $\Lambda$ and $\Sigma$ baryons, but in a color octet
state, coupled to a gluino \re{bucella}).  There are two R-baryon
flavor octets with $J=1$.  Neglecting the mixing between the states,
the states with quarks contributing spin $3/2$ have masses of about
$385-460 \MeV$ above that of the $S^0$ and the states with quarks
contributing spin $1/2$ have masses of about $815-890 \MeV$ above that
of the $S^0$.  If we require that the photino is a significant dark
matter component so $1.3 \leq M_{\r0}/m_{\pho} \leq 1.6$ according to
Ref.\ \re{cfk}, and take the mass of $\r0$ to be about $1.6-1.8 \GeV$
as expected, then $m_\pho$ lies in the range $0.9 \sim 1.3\GeV$.  If
we assume that $S^0$ is minimally stable, we have $m_{S^0} \approx m_p
+ m_\pho$ resulting in $m_{S^0}$ in the range $1.9$ to $2.3\GeV$.  The
other resonance parameters are fixed at the same values as those for
the protons.

In Fig.\ 1, we show the proton energy and the $S^0$ energy today (with
$h=0.5$) if it had been injected at a redshift $z$ (or equivalently
from the corresponding distance\footnote{Marked are the
luminosity distances $d_L=H_0^{-1} q_0^{-2}
\left[ z q_0 + (q_0-1) ( \sqrt{2q_0 z +1} -1) \right]$ where the
deceleration parameter $q_0$ is 1/2 in our $\Omega_0=1$ universe.})
with an energy of $10^{22}\eV$, $10^{21}\eV$, and $10^{20}\eV$.  To
explore the interesting mass range, we have set the $S^0$ mass to  $1.9
\GeV$ in the upper plot while we have set it to $2.3 \GeV$ in the lower
plot.  For the cosmic rays arriving with $10^{20}\eV$, the distance
is increased by more than thirty times, while for those arriving with
$10^{19.5}\eV$, the distance is increased by about fifteen
times.   In Fig.\ 2, we recalculate the energies with $h=0.8$.

Using the mean energy approximation, we can also calculate the evolved
spectrum of the primary $S^0$ spectrum observed on Earth given the
initial spectrum at the source (where all the particles are injected
at one time).  With the source at $z=0.54$ (the source distance for 3C
147) and the initial spectrum having a power law behavior of $E^{-2}$,
the evolved spectrum is shown in Fig.\ 3.  We see that even though
there is significant attenuation for the $S^0$ number at $3 \times
10^{20}\eV$ for most of the cases shown, when the overall cross
section (which was originally estimated quite conservatively) is
reduced by a factor of half, the bump lies very close to the Fly's Eye
event.  Moreover, taking the Fly Eye's event energy to be $2.3 \times
10^{20}\eV$ which is within $1 \sigma$ error range, we see that the
$S^0$ can easily account for the Fly Eye's event.  For sources such as
MCG 8-11-11, $S^0$ clearly can account for the observed event without
upsetting the proton flux at lower energies.

\begin{figure}[t]
\hspace*{25pt} \epsfxsize=400pt \epsfbox{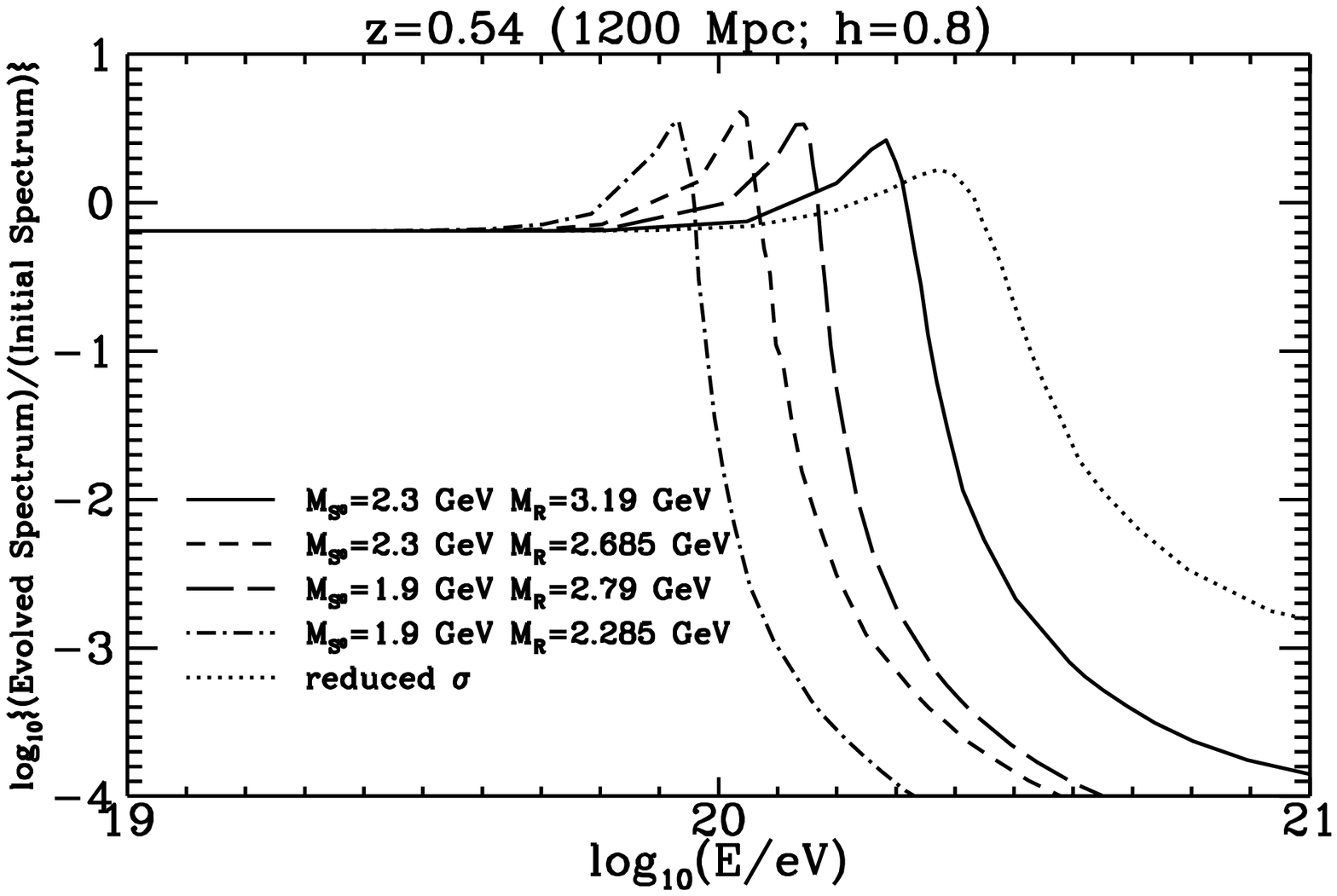}\\
\hspace*{1em} {\footnotesize{{\bf Fig.\ 3:} An initial $S^0$ injection
spectrum having a power law form of $E^{-2}$ is evolved through the
particle's interaction with the CBR during its 1200 Mpc travel to
Earth.  The masses of the $S^0$ and its associated resonance are
shown.  The curve labeled reduced $\sigma$ has the same mass
parameters as the solid curve except with our conservative estimation
of the total cross section reduced by a factor of half.}}
\end{figure}

\vspace{36pt}
\centerline{\bf IV.  CONCLUSIONS}
\vspace{24pt}

We have considered the suggestion that the very long-lived or stable
new hadron called $S^0$, a $uds$-gluino bound state predicted 
in some supersymmetric models, can account for the primary cosmic ray
particles at energies above the GZK cutoff.  We noted ways that conventional
acceleration mechanisms might result in acceptable fluxes of high energy
$S^0$'s.  We also found that the $S^0$ can propagate at least fifteen to
thirty times longer through the CBR than do nucleons, for the same amount
of total energy loss.  Thus, if $S^0$'s exist and there exists an acceleration
mechanism which can generate an adequate high-energy spectrum, $S^0$'s can
serve as messengers of the phenomena which produce them, allowing MCG 8-11-11
Seyfert galaxy or 3C 147 quasar to be viable sources for these ultrahigh
energy cosmic rays.   

Although much of the relevant hadronic physics in the atmospheric
shower development will be similar to that for the proton primaries,
some subtle signatures of an $S^0$ primary are still expected.  Because
an $S^0$ is expected to have a cross section on nucleons or
nuclei somewhere between $1/10$ and
$4/3$ of the $p$-$p$ cross section, the depth of the shower
maximum may be a bit larger than that due to the proton.  Furthermore,
because it is about twice as massive as the proton, it deposits its
energy a bit more slowly than a proton, broadening the distribution of
the shower.  There may be further signatures in the shower
development associated with the different branching fractions to
mesons, but we leave that numerical study for the future.

A prediction of this scenario which can be investigated after a large
number of UHE events have been accumulated is that UHE cosmic rays
primaries point to their sources.  If there are a limited number of
sources, multiple UHE events should come from the same direction.
Also, the UHE cosmic-ray spectrum from each source should exhibit a
distinct energy dependence with a cutoff (larger than the GZK cutoff)
at an energy which depends on the distance to the source.  The
systematics of the spectrum in principle could reveal information
about both masses of supersymmetric particles and the primary spectrum of
the source accelerator.  

We noted that the mass range for a new hadron which can account for the
observed properties of UHE cosmic ray events is limited:  it must be
at least 2 GeV in order to evade the GZK bound, yet small enough that
the atmospheric shower it produces will mimic an ordinary hadronic shower.

\vspace{36pt}
\centerline{\bf ACKNOWLEDGMENTS}
\vspace{24pt}
DJHC thanks Angela Olinto and Ted Quinn for useful discussions about aspects
of UHE cosmic rays.  We thank D. Goulianos and P. Schlein for
discussions about Regge lore and high energy $p-p$ scattering, and
A. Riotto for discussions about extra colored particles and alternate
candidates for the uhecron.  DJHC and EWK were supported by the DOE
and NASA under Grant NAG5-2788.  GRF was supported by the NSF
(NSF-PHY-94-2302).  

\frenchspacing
\def\prpts#1#2#3{Phys. Reports {\bf #1}, #2 (#3)}
\def\prl#1#2#3{Phys. Rev. Lett. {\bf #1}, #2 (#3)}
\def\prd#1#2#3{Phys. Rev. D {\bf #1}, #2 (#3)}
\def\plb#1#2#3{Phys. Lett. {\bf #1B}, #2 (#3)}
\def\npb#1#2#3{Nucl. Phys. {\bf B#1}, #2 (#3)}
\def\apj#1#2#3{Astrophys. J. {\bf #1}, #2 (#3)}
\def\apjl#1#2#3{Astrophys. J. Lett. {\bf #1}, #2 (#3)}
\begin{picture}(400,50)(0,0)
\put (50,0){\line(350,0){300}}
\end{picture}

\vspace{0.25in}

\def\labelenumi{[\theenumi]}

\begin{enumerate}

\item \label{measure} J.\ Linsley, \prl{10}{146}{1963}; N.\ Hayashida
et al., \prl{73}{3491}{1994}; D. J. Bird et al., \apj{424}{491}{1994}.

\item\label{flyseye}  D. J. Bird et al., \apj{441}{144}{1995}.

\item\label{bierman} P. L. Bierman, J. Phys. G: Nucl. Part. Phys. 
{\bf 23}, 1 (1997).

\item\label{halzen} F. Halzen et al., Astropart. Phys. {\bf 3}, 151 (1995).  

\item \label{elbert} J.\ W.\ Elbert and P.\ Sommers,
\apj{441}{151}{1995}.

\item\label{burdman} G. Burdman et al., hep-ph/9709399.  

\item\label{gf96} G. R. Farrar, \prl{53}{4111}{1996}.

\item\label{gf97}  G. R. Farrar, hep-ph/971077.

\item\label{farrarprd95} G.\ Farrar, \prd{51}{3904}{1995}.

\item\label{afk} I. F. Albuquerque, G. R. Farrar, and E. W. Kolb, in
preparation. 

\item\label{hs} C.\ T.\ Hill and D.\ N.\ Schramm, \prd{31}{564}{85}.

\item \label{gqw} J.\ Geddes, T.\ Quinn, and  R.\ Wald, astro-ph/9506009.

\item\label{albuquerque} I. Albuquerque et al., \prl{78}{3252}{1997}.

\item \label{rachen} J. P. Rachen and P. L. Biermann, Astron. 
Astrophys. {\bf 272}, 161 (1993).

\item\label{onepush} K. S. Cheng, C. Ho, and M. Ruderman,
\apj{300}{500}{1986}; Ref.\ [3].

\item\label{kr} V. A. Kuzmin and V. A. Rubakov, astro-ph/9709178.

\item\label{berezinsky} V.\ Berezinsky and M. Kachelreiss, hep-ph/9709485.

\item\label{plaga} R.\ Plaga, \prd{51}{6504}{1995}.

\item\label{jwc} J. W. Cronin, Nucl. Phys. Proc.\ Suppl.\ {\bf 28B}, 234 (1992).

\item \label{crossdata} {\em Total Cross-sections for Reactions of
High Energy Particles}, ed. H. Schopper \\ (Springer Verlag, New York, 1988).

\item \label{jost} R.\ Jost, J.\ M.\ Luttinger, and M.\ Slotnick,
\prd{80}{189}{1950}.

\item\label{bucella} F. Bucella, G. R. Farrar, A. Pugliese, 
\plb{153}{311}{1985}.

\item \label{cfk} D. J. H. Chung, G. R. Farrar, and E. W. Kolb,
\prd{56}{6096}{1997}.

\end{enumerate}

\end{document}